\pdfoutput=1
\documentclass[11pt]{article}

\usepackage{amsmath,amssymb}
\usepackage{dsfont}        
\usepackage{empheq}        
\usepackage{mathtools}     
\usepackage{dcolumn}
\usepackage{bm}
\usepackage{slashed}
\usepackage{graphicx}
\usepackage{authblk}
\usepackage{ulem}
\usepackage{CJK}
\usepackage{tikz}
\usetikzlibrary{decorations.pathmorphing}
\graphicspath{{figures/}}
\usepackage{hyperref}
\hypersetup{hidelinks}
\usepackage{xcolor}

\newcommand{\rev}[1]{{#1}}
\newenvironment{revision}{\begingroup}{\endgroup}
\definecolor{revisionpurple}{HTML}{4B176D}
\newcommand{\vfourrev}[1]{{#1}}
\makeatletter
\let\original@cite\@cite
\def\@cite#1#2{{\color{black}\original@cite{#1}{#2}}}
\makeatother

\def\L{{\Lambda}}
\def\d{{\delta}}

\def\a{{\alpha}}
\def\b{{\beta}}
\def\c{{\chi}}
\def\f{{\phi}}
\def\g{{\gamma}}

\def\h{\eta}

\def\P{{\Pi}}
\def\m{{\mu}}
\def\n{{\nu}}
\def\r{{\rho}}
\def\s{{\sigma}}

\def\t{{\tau}}

\def\ps{{\psi}}

\def\P{{\Pi}}

\def\({\left(}
\def\){\right)}
\def\[{\left[}
\def\]{\right]}
\def\jp{{J/\ps}}

\newcommand{\dg}{\dagger}

\newcommand{\tr}{\text{tr}}

\date{\today}

\begin{document}

\begin{CJK}{UTF8}{gbsn}

\title{\bf Anisotropic chromo-field fluctuations and spin alignment of quarkonia}

\author[1]{{Yuhao Liang}
\thanks{liangyh83@mail2.sysu.edu.cn}}
\affil[1]{School of Physics and Astronomy, Sun Yat-sen University, Zhuhai 519082, China}
\author[1]{{Shu Lin}
\thanks{linshu8@mail.sysu.edu.cn}}

\maketitle

\begin{abstract}

 Spin correlations are believed to contribute significantly to vector-meson spin alignment. In this work, we study the generation of spin correlations through fluctuations of the chromomagnetic field in the quark-gluon plasma. We show that anisotropic fluctuations generically lead to finite vector-meson spin alignment. We study two sources of anisotropic chromomagnetic-field fluctuations: one arises from the motion of a vector meson relative to the medium, which makes isotropic fluctuations in the medium anisotropic in the meson rest frame; the other arises from QGP shear flow through viscous corrections to the chromomagnetic field fluctuations. The linear HTL estimates indicate sensitivity to the nonperturbative magnetic sector; we treat the full field correlators as phenomenological inputs.We apply the mechanism to study spin alignment of $J/\psi$ in heavy ion collisions. With QGP evolution modeled by Bjorken flow, we obtain negative spin alignment from both sources, with the shear induced contribution numerically suppressed compared to the motion induced contribution.

\end{abstract}

\newpage

\section{Introduction}

The observation of global spin polarization of $\Lambda$ hyperons in heavy-ion collisions (HICs) has established spin as a valuable probe of the vortical structure of the quark-gluon plasma (QGP) \cite{STAR:2017ckg,STAR:2019erd}. Measurements by the STAR Collaboration have shown that $\Lambda$ and $\bar{\Lambda}$ hyperons acquire a finite polarization along the direction of the system's global angular momentum, indicating that the hot QCD medium can transfer orbital angular momentum to particle spin \cite{Becattini:2013fla,Fang:2016vpj}. At the microscopic level, the polarization of $\Lambda$ hyperons is understood to originate from the polarization of strange quarks during the evolution of the medium and their subsequent hadronization \cite{Liang:2004ph}.

The same quark polarization mechanism is expected to induce spin alignment of vector mesons, such as $\phi$ and $J/\psi$, through the recombination of polarized quark-antiquark pairs \cite{Liang:2004xn,Yang:2017sdk}. Since vector-meson spin alignment is generated by the correlation between the spins of the constituent quark and antiquark, the resulting contribution is of second order in the quark polarization. Quantitative estimates based on this mechanism generally predict a spin alignment much smaller than that observed experimentally for $\phi$ and $J/\psi$ mesons in heavy-ion collisions \cite{ALICE:2019aid,STAR:2022fan,ALICE:2022dyy}. This discrepancy suggests that additional sources of spin alignment independent of individual quark polarization may exist. Various mechanisms of spin alignment have been explored in the literature \cite{Sheng:2022wsy,Sheng:2023urn,Kumar:2023ghs,Yang:2024qpy,Li:2022vmb,Wagner:2022gza,DeMoura:2023jzz,Xu:2024kdh,Sheng:2024kgg,Fu:2023qht,Zhao:2024ipr,Chen:2024hki,Chen:2025mrf,Liang:2025hxw,Yan:2025tlx,Zhu:2025rdj}. Anisotropic curvature of the freeze-out hypersurface has also been proposed as a geometric source of vector-meson tensor polarization~\cite{Zhang:2025freezeout}.

The importance of spin correlation has been emphasized in recent years \cite{Sheng:2022wsy,Sheng:2023urn,Kumar:2023ghs,Yang:2024qpy,Lv:2024uev,Oliva:2026wbo}. Experimental measurements of spin correlation in $\Lambda$-hyperon pairs produced in proton-proton collisions have revealed nontrivial spin-correlation effects, which can be traced back to the creation of correlated $s\bar{s}$ pairs from the QCD vacuum \cite{STAR:2025njp}. These observations demonstrate that spin correlations are naturally generated in strong-interaction processes and may survive hadronization. In order to generate vector-meson spin alignment with respect to the event plane in heavy-ion collisions, however, the relevant spin correlations must be produced inside the medium and be correlated with the geometry or collective dynamics of the collision.

Several mechanisms for such in-medium spin correlations have been proposed. One class of mechanisms is based on fluctuations of strong force fields in the hadronic phase \cite{Sheng:2022wsy,Sheng:2023urn}. Significant strong force field fluctuation is needed to understand the measured spin alignment, indicating that hadronic interactions may contribute substantially to vector-meson spin alignment. Another class of mechanisms is based on fluctuations of chromo-magnetic fields in the deconfined phase \cite{Kumar:2023ghs,Yang:2024qpy}. In particular, spin correlations generated in the early-time Glasma stage, characterized by strong coherent color fields, have been extensively discussed in the literature. Quarkonium polarization in the QGP has also been studied using an open quantum system framework with gauge-invariant chromomagnetic correlators~\cite{Yang:2024ejk}. A gauge-covariant formulation of the coupled spin and color dynamics of massive particles in a background Yang--Mills field has been developed in Ref.~\cite{Zhou:2025spin}. These studies highlight the potential importance of gluonic field fluctuations as a source of spin alignment.

In this work, we investigate in-medium spin correlations generated in the QGP phase, which spans long time and is thus expected to play an important role in the building up of spin correlation. A key observation is that a nonvanishing contribution to vector-meson spin alignment requires anisotropic chromo-magnetic field fluctuations. If the fluctuations are completely isotropic, the induced spin correlations do not generate a preferred spin alignment with respect to the event plane. We explore two possible mechanisms that can generate such anisotropy in the QGP.
The first mechanism arises from the motion of the vector meson relative to the  QGP. Even if the chromo-magnetic field fluctuations are isotropic in the fluid rest frame, a moving vector meson experiences a Lorentz-transformed field configuration that becomes anisotropic in its rest frame. This kinematic effect can therefore induce direction-dependent spin correlations. The second mechanism originates from the shear flow of the QGP. Velocity gradients in the medium modify the structure of chromo-magnetic field fluctuations and naturally generate anisotropy in the fluctuation spectrum. Since the shear flow is correlated with the collective expansion geometry of the fireball, the resulting spin correlations are also correlated with the event plane and can contribute to the observed vector-meson spin alignment.
The mechanisms of generating spin alignment from anisotropic
chromo-magnetic fluctuations are applicable for generic vector meson
interacting with the QGP. To be specific, we apply
the mechanisms to $\jp$. We shall quantify the
spin correlation from the two mechanisms above and assess their contributions to the $\jp$ spin alignment measured in heavy-ion collisions.

The remainder of this paper is organized as follows. In Sec.~\ref{sec:density-matrix}, we derive the spin correlation in a color-singlet quarkonium state from the chromomagnetic-spin interaction, which is used to generate spin alignment. In Sec.~\ref{sec:anisotropy}, we compute the anisotropic fluctuations of the chromomagnetic field from the motion of quarkonia and from shear stress of the QGP. In Sec.~\ref{sec:jpsi}, we evaluate the spin alignment of $\jp$ produced at LHC energy arising from these fluctuation sources. We conclude and discuss future directions in Sec.~\ref{sec_outlook}.

\section{Spin alignment of quarkonia from anisotropic chromomagnetic field fluctuations}
\label{sec:density-matrix}

The general relation between spin alignment and spin correlations has been derived in \cite{Lv:2024uev}. Such correlations must be generated dynamically. In the deconfined phase, they can arise from the spin-chromomagnetic-field interaction. Spin correlations generated by this mechanism have been studied in the Glasma phase in \cite{Kumar:2023ghs,Yang:2024qpy}. In this section, we study the corresponding mechanism in the QGP phase. A distinguishing feature of the QGP phase is its long lifetime that allows spin to respond to chromomagnetic field.
Like ordinary magnetic field, chromomagnetic field can polarize a charm quark. In QGP, expectation value of chromomagnetic field vanishes by color neutrality. Nevertheless nonvanishing fluctuation of chromomagnetic field can lead to spin correlations of a $Q\bar{Q}$ pair. If the fluctuating field is anisotropic, i.e. has a preferred direction, it can contribute to spin alignment of the quarkonia. In this picture, the anisotropic chromomagnetic field fluctuations can be viewed as source of spin alignment. 

Three time scales are involved in the dynamics of spin alignment. Spin relaxation time $\t_s$, correlation time of chromomagnetic field $\t_B$ and evolution time of QGP $\t_{QGP}$. $\t_s$ measures the response of spin to chromomagnetic field and is found to be parametrically large in the heavy quark limit $\t_s\sim M_Q^2/T^3$ \cite{Hongo:2022izs}. On dimensional ground, we expect generically $\t_B\sim 1/T$. We shall further assume that $\t_{QGP}\gg \t_s$. Thus the full hierarchy in time scales are $\t_{QGP}\gg \t_s\gg \t_B$. %

The second assumption concerns the hierarchy between the correlation length of the chromomagnetic field $l_B$ and the quarkonium size $r_{Q\bar{Q}}$. Generically we expect $l_B\sim 1/T$. The size of $\jp$ is assumed to be much smaller than $l_B$, so that the quarkonium will be treated as point-like. %

We are ready to derive the spin density matrix of the color-singlet
quarkonium state. The spin-chromomagnetic interaction is given by $\d{\cal L}=\int dt\,{\bf S}_a(t)\cdot{\bf B}_a(t,\bar{x})$. ${\bf S}_a$ is the chromomagnetic moment for quark defined as
\begin{equation}
{\bf S}_a(t)=\int d^3x\frac{g_s}{2M_Q}{\f}^\dg(t,x){\boldsymbol{\sigma}}T_a\f(t,x)
\label{eq:chromomagnetic-moment}
\end{equation}
where $\f(t,x)$ is the large component of the quark spinor. ${\bf S}_a(t)$ will be loosely referred to as quark spin. The counterpart for antiquark is obtained by the substitution $T^a\to-T^a{}^*$.
Since we have taken the quarkonium to be point-like, the chromomagnetic field is to be taken at the location of the quarkonium $\bar{x}$, which should be understood as a coarse-grained coordinate. 
The spin-chromomagnetic interaction gives the following retarded response of quark spin density to chromomagnetic field
\begin{align}
S_a^i(t)&=i\int dt'\theta(t-t')\langle[S_a^i(t),S_c^k(t')B_c^k(t',\bar{x})]{\rangle}\nonumber\\
&=\int dt'G_{s,ac}^{ik}(t-t')B_c^k(t',\bar{x}),
\label{eq:TAG-spin-response}
\end{align}
with $G_{s,ac}^{ik}(t-t')=i\theta(t-t'){\langle}[S_a^i(t),S_c^k(t')]{\rangle}$ being the retarded spin {response} function.
A similar expression holds for response of anti-quark spin ${S}_b^j(t)$. The spin correlation is generated from fluctuation of chromomagnetic field as
\begin{align}
\langle S_a^i(t){S}_b^j(t)\rangle=-\int dt'dt''G^{ik}_{s,ac}(t-t')G^{jl}_{s,bd}(t-t'')\langle B_c^k(t',\bar{x})B_d^l(t'',\bar{x})\rangle.
\label{eq:TAG-induced-correlation}
\end{align}
The minus sign is taken from the coupling $-T^a{}^*$ in the chromomagnetic moment for antiquark.
We adopt a gauge-covariant model for spin relaxational dynamics,
\begin{equation}
\left(D_t^{ab}+\frac{\delta^{ab}}{\tau_s}\right)S_b^i=0,
\qquad D_t^{ab}=\delta^{ab}\partial_t-ig_sA_0^c(t,\bar x)(T_A^c)_{ab}.
\label{eq:covariant-relaxation}
\end{equation}
The corresponding retarded response is
\begin{equation}
G_{s,ab}^{ij}(t,t')=\frac{g_s}{2M_Q}\frac{\chi_s}{\tau_s}
e^{-(t-t')/\tau_s}[U_A(t,t')]_{ab}\delta^{ij}\theta(t-t'),
\label{eq:covariant-spin-response}
\end{equation}
with $U_A(t,t')$ being the adjoint temporal Wilson line
	\begin{equation}
	U_A(t,t')={\cal P}\exp\left[ig_s\int_{t'}^t d\lambda\,
	A_0^c(\lambda,\bar x)T_A^c\right],
	\qquad (T_A^c)_{ab}=-if^{cab}.
	\label{eq:adjoint-worldline-link}
	\end{equation}
The parameters in Eq.~\eqref{eq:covariant-spin-response} will be explained shortly.
It is instructive to absorb the Wilson line in the fluctuation of chromomagnetic field as
\begin{align}
{\cal C}_{B,ab}^{kl}(t)=\langle [U_A(t,t')]_{ac}B_c^k(t')[U_A(t,t'')]_{bd}B_d^l(t'')\rangle.
\label{eq:TAG-field-spectrum-U}
\end{align}
Using the relation $[U_A(t,t')]_{ba}T^b=U(t,t')T^aU(t',t)$ with $T^a$ and $U$ being respectively generator and Wilson line in the fundamental representation, we can show \eqref{eq:TAG-field-spectrum-U} transforms as
\begin{align}\label{transformation}
{\cal C}_{B,ab}^{kl}(t)T^aT^b\to V(t){\cal C}_{B,ab}^{kl}(t)T^aT^bV^\dagger(t), 
\end{align}
with $V(t)$ being the gauge transformation matrix. Note that $Tr[T^aT^b]=\frac{1}{2}\delta^{ab}$, which corresponds to color average of the chromomagnetic field fluctuation. From \eqref{transformation}, it follows that the color average fluctuation is gauge invariant, which motivates the following gauge invariant definition of the chromomagentic fluctuation as
\begin{equation}
{\cal C}_B^{ij}(t,\bar{x})=\frac{1}{d_A}\delta^{ab}\langle [U_A(t,t')]_{ac}B_c^{\vfourrev{i}}(t')[U_A(t,t'')]_{bd}B_d^{\vfourrev{j}}(t'')\rangle,
\label{inv_fluc}
\end{equation}
with $d_A=N_c^2-1$. \vfourrev{Field-fluctuation correlators are understood to be symmetrized throughout.}
It follows that the following color averaged spin correlation is also gauge invariant
\begin{align}
\frac{1}{d_A}\delta^{ab}\langle S_a^i(t)S_b^j(t)\rangle
\end{align}

\begin{figure}
  \centering
  \begin{tikzpicture}[x=1cm, y=1cm,
    heavy/.style={line width=0.75pt},
    gluon/.style={decorate, decoration={coil, aspect=0.45, segment length=4.5pt, amplitude=2.6pt}, line width=0.75pt}]
    \draw[heavy] (1,3.2) -- (5,3.2);
    \draw[heavy] (1,2.0) -- (5,2.0);
    \draw[gluon] (3,3.2) -- (3,2.0);
    \node[above] at (3,3.2) {$T^a_{\alpha\beta}$};
    \node[below] at (3,2.0) {$(T^b_{\delta\gamma})^*$};
    \node[above] at (1.8,3.2) {$\beta$};
    \node[above] at (4.2,3.2) {$\alpha$};
    \node[below] at (1.8,2.0) {$\gamma$};
    \node[below] at (4.2,2.0) {$\delta$};
    \node[left]  at (1,2.6) {$\delta_{\beta\gamma}$};
    \node[right] at (5,2.6) {$\delta_{\alpha\delta}$};
  \end{tikzpicture}
    \caption{{Chromomagnetic exchange between the heavy quark
  $Q$ (upper line) and antiquark $\bar Q$ (lower line). The color indices are
  shown explicitly: the quark line carries $\beta\to\alpha$ with the vertex
  factor $T^a_{\alpha\beta}$, and the antiquark line carries
  $\gamma\to\delta$ with the vertex factor $(T^b_{\delta\gamma})^*$. The
  initial and final quarkonium are color singlets, giving the contractions
  $\delta_{\beta\gamma}$ and $\delta_{\alpha\delta}$ on the left and right,
  respectively. Together they reproduce
  $T^a_{\alpha\beta}(T^b_{\delta\gamma})^*\delta_{\beta\gamma}\delta_{\alpha\delta}
  =\tr(T^aT^b)$ in Eq.~\eqref{eq:color-factor}.}}
\label{fig:diagrams}
\end{figure}
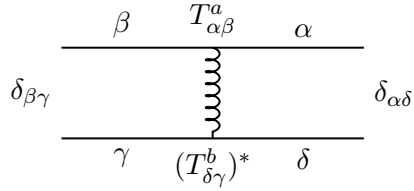

We proceed in the temporal axial gauge (TAG) $A^0=0$, which leads to $U_A=\mathbf1$. It is then convenient to perform Fourier transform of Eq.~\eqref{eq:covariant-spin-response} to give

\begin{align}
G_{s}^{ik}(q_0)=\frac{g_s}{2M_Q}\frac{\c_s}{1-iq_0\t_s}\d^{ik},
\label{eq:TAG-frequency-response}
\end{align}
with $\frac{g_s}{2M_Q}\c_s$ the static susceptibility of the chromomagnetic moment. We have factored out the trivial color factor in TAG. With the hierarchy $\t_{QGP}\gg\t_s$, we obtain
\begin{align}
\frac{1}{d_A}\delta^{ab}\langle S_a^i(t)S_b^j(t)\rangle=-\int_{q_0,q}G^{ik}_s(q_0)G^{jl}_s(-q_0){\cal C}_B^{kl}(q_0,q),
\end{align}
with $\int_{q_0}=\int\frac{dq_0}{2\pi}$ and $\int_q=\vfourrev{\int}\frac{d^3q}{(2\pi)^3}$.
The hierarchy $\t_s\gg\t_B$ implies that the slow spin relaxation effectively probes only the low frequency limit of $\vfourrev{{\cal C}_B^{kl}}$ as the response function $\vfourrev{G_s^{ik}(q_0)}$ is strongly suppressed at the scale of $q_0\sim\t_B^{-1}$. Thus we may approximate the chromomagnetic fluctuation with its static limit $\vfourrev{{\cal C}_B^{kl}(q_0)\simeq{\cal C}_B^{kl}(0)}$. We can then perform the $q_0$-integration to obtain
\begin{align}
\frac{1}{d_A}\delta^{ab}\langle S_a^i(t){S}_b^j(t)\rangle&\simeq-\left(\frac{g_s}{2M_Q}\right)^2\frac{\c_s^2}{2\t_s}\int_q{\cal C}_{B}^{ij}(0,q).
\label{eq:spin-correlation-static}
\end{align}
Finally we can convert the chromomagnetic moment to spin for color singlet state as
\begin{align}
\langle S_a^i(t){S}_b^j(t)\rangle=\left(\frac{g_s}{2M_Q}\right)^2C\delta_{ab}\langle S^i(t)S^j(t)\rangle.
\label{eq:color-singlet-spin-correlation}
\end{align}
The identity is to be understood with expectation value taken on a color singlet state. Clearly the spin correlation defined this way is gauge invariant. Referring to Fig.~\ref{fig:diagrams}, we obtain for the color factor on the left hand side the following
\begin{align}
&T^a_{\a\b}\(T^b_{\d\g}\)^*\d_{\b\g}\d_{\a\d}=T^a_{\a\b}T^b_{\g\d}\d_{\b\g}\d_{\a\d}=\tr(T^aT^b)=\frac{1}{2}\d_{ab}.%
\label{eq:color-factor}
\end{align}
For the right side, we simply substitute the color generators by identity to have $\d_{\a\b}\d_{\d\g}\d_{\b\g}\d_{\a\d}=N_c$. It gives then $C=\frac{1}{2N_c}$. We can then rewrite \eqref{eq:spin-correlation-static} as
\begin{align}\label{spin_correlation}
\langle S^i(t){S}^j(t)\rangle&\simeq-\frac{\c_s^2}{2\t_sC}\int_{q}{\cal C}^{ij}_{B}(0,q),
\end{align}
We have expressed the spin correlation as a response to quadratic fluctuation of chromomagnetic field. It is possible to introduce a spin precession term in \eqref{eq:covariant-relaxation}, which would result in additional response to quatic fluctuation. We restrict to quadratic fluctuation for simplicity in this paper.

It remains to relate spin correlation to spin alignment, which is worked out in Appendix~\ref{sec_app_general_density}. Identifying $\langle S^iS^j\rangle$ with $\langle P^i\bar{P}^j\rangle$ in \eqref{eq:app-rho00-structure}, we have
\begin{align}
  \rho_{00}-\frac{1}{3}
  &=\frac{1}{3}\langle S^iS^j\rangle
  \left(\frac{2}{3}\d_{ij}-2n_in_j\right)
  +O(P^4),
\label{eq:spin-alignment-proportional}
\end{align}
where ${\bf n}$ is the quantization axis.
In the following, we retain the leading $O(P^2)$ term and neglect higher order corrections.

\section{Anisotropic chromomagnetic field fluctuations}
\label{sec:anisotropy}

In Sec.~\ref{sec:density-matrix}, we showed that the spin correlation of a color-singlet, spin-triplet quarkonium state is determined by the fluctuation of the chromomagnetic field,
$\langle B_i^aB_j^a\rangle$. If the fluctuation is
isotropic, $\langle B_i^aB_j^a\rangle\propto\delta_{ij}$, the three spin states are populated equally and no spin alignment is generated.
A non-vanishing spin alignment therefore directly probes the
anisotropy of chromomagnetic fluctuations in the medium.
In equilibrium, chromomagnetic-field fluctuations are isotropic by rotational invariance. Anisotropic fluctuations require an off-equilibrium effect. We consider two such effects: quarkonium motion relative to the QGP and QGP shear flow. The former is an effective off-equilibrium effect as it is from a moving equilibrium medium seen by the quarkonium. The latter is characterized by a hydrodynamic gradient. We focus on shear because it is the tensor component of the velocity gradient and can therefore induce the tensor component of the spin density matrix at linear order. Vector-type hydrodynamic gradients contribute to spin alignment only at quadratic order. Momentum-space anisotropy of the QGP associated with its rapid longitudinal expansion also renders the interquark potential anisotropic and modifies quarkonium binding~\cite{Dong:2021Debye}; such medium-induced modifications of quarkonium binding are beyond the scope of this work.

In Sec.~\ref{sec:chap3:general}, we work out fluctuations of chromomagnetic field in equilibrium. We also study fluctuation of chromoelectric field, which will turn into part of the chromomagnetic field seen by the moving quarkonium. In
Sec.~\ref{sec:chap3:relative-motion} and Sec.~\ref{sec:chap3:shear}, we derive the anisotropic chromomagnetic fluctuation induced by the relative motion and shear of the QGP respectively. The HTL estimates indicate sensitivity to the nonperturbative magnetic sector.

\subsection{Chromo field fluctuations in equilibrium}
\label{sec:chap3:general}

From the discussions in Sec.~\ref{sec:density-matrix}, the quantity of our interest is $\vfourrev{\int_q{\cal C}_B^{ij}(0,q)}$. In equilibrium, we can parameterize
\begin{align}
\rev{\int_q{\cal C}_{B}^{ij}(0,q)
=\d^{ij}BB_\text{stat}.}
\end{align}
\rev{The hard thermal loop (HTL) calculation with linear field strengths in TAG, presented in Appendix~\ref{sec_app_equilibrium_htl}, gives logarithmic infrared-cutoff dependence for the zero-frequency magnetic correlator. At the soft scale $k\sim g_sT$, preliminary power counting suggests that nonlinear field-strength contributions in this HTL estimate are suppressed by an additional power of $g_s$; see Appendix~\ref{sec_app_equilibrium_htl}. This infrared sensitivity motivates treating $BB_\text{stat}$ as a phenomenological input sensitive to the nonperturbative magnetic sector of hot QCD~\cite{Bellac:2011kqa}. The ultraviolet sensitivity is regulated by a cutoff of order $r_{Q\bar{Q}}^{-1}$ in the point-like model.} We will also need chromoelectric analog of the above, defined as
\begin{align}
\rev{\int_q{\cal C}_{E}^{ij}(0,q)
=\d^{ij}EE_\text{stat}.}
\end{align}
\rev{The linear-field-strength HTL electric contribution evaluated in Appendix~\ref{sec_app_equilibrium_htl} is infrared finite because of Debye screening.}

While isotropic fluctuations in equilibrium do not contribute to spin alignment, the combination $EE_\text{stat}+BB_\text{stat}$ does contribute to spin alignment of a moving quarkonium, within the local-field estimate used in Sec.~\ref{sec:chap3:relative-motion}. We treat $BB_\text{stat}$ and $EE_\text{stat}$ as phenomenological inputs and parameterize their contribution by
\begin{equation}
 \frac{\chi_s^2}{2\tau_s}
 \bigl(BB_\text{stat}+EE_\text{stat}\bigr)
 \equiv C\,\Xi_{\rm eq}.
\label{eq:equilibrium-response-input}
\end{equation}
The single dimensionless parameter $\Xi_{\rm eq}$ contains the spin
susceptibility, the relaxation time, and static fluctuations.   The color factor $C$ is inserted for convenience.  Taking $g\simeq 2$ for realistic
coupling and $T\sim \L_\text{QCD}$ at freezeout, with
$\c_s\sim r_{Q\bar{Q}}^{2}$\footnote{The susceptibility is defined inside the quarkonium state, so it is a function of the quarkonium structure rather than the temperature.}, $\t_s\sim M_Q^2/T^3$, and
$BB_\text{stat}+EE_\text{stat}\sim T^3$, gives
$\Xi_{\rm eq}\sim r_{Q\bar{Q}}^4T^6/M_Q^2$.

\subsection{Anisotropy from quarkonium motion in QGP}
\label{sec:chap3:relative-motion}

For a moving quarkonium in QGP, it observes Lorentz-boosted chromo fields.
Let $\vec p_{\rm cm}$ be the quarkonium three-momentum in the local rest
frame of the QGP, with
$p_{\rm cm}=|\vec p_{\rm cm}|$ and
$\hat p_{\rm cm}=\vec p_{\rm cm}/p_{\rm cm}$. The boost direction $\hat p_{\rm cm}$ then supplies the only preferred axis. The corresponding Lorentz
factor and velocity are
\begin{equation}
\gamma
=
\frac{E_{Q\bar Q}}{2M_Q}
=
\frac{\sqrt{4M_Q^2+p_{\rm cm}^2}}{2M_Q},
\qquad
\vec v=\frac{\vec p_{\rm cm}}{E_{Q\bar Q}}.
\end{equation}
The chromomagnetic field in the quarkonium rest frame is given by
\begin{equation}
B'_i
=
(1-\gamma)(\hat p_{\rm cm})_i(\hat p_{\rm cm})_jB_j
+\gamma B_i
-\gamma v
\varepsilon_{ijk}(\hat p_{\rm cm})_jE_k .
\label{eq:Bprime}
\end{equation}
Assuming isotropic fluctuation for (unprimed) chromo fields in QGP frame, the two projections of chromomagnetic fluctuations entering \eqref{eq:spin-alignment-proportional} are
\begin{align}
\bigl\langle(B')_{stat}^2\bigr\rangle
&=
BB_{stat}\left(3+\frac{p_{\rm cm}^2}{2M_Q^2}\right)
+EE_{stat}\frac{p_{\rm cm}^2}{2M_Q^2},
\label{eq:Bprime-sq}
\\[4pt]
\bigl\langle(\hat n\!\cdot\!\vec B')_{stat}^2\bigr\rangle
&=
BB_{stat}\left[
1+\frac{p_{\rm cm}^2}{4M_Q^2}
-\frac{p_{\rm cm}^2}{4M_Q^2}
(\hat n\!\cdot\!\hat p_{\rm cm})^2
\right]
\notag\\
&\quad
+EE_{stat}\frac{p_{\rm cm}^2}{4M_Q^2}
\left[1-(\hat n\!\cdot\!\hat p_{\rm cm})^2\right].
\label{eq:nBprime-sq}
\end{align}
The first combination is independent of
the spin-quantization direction, whereas the second carries the
quadrupolar angular dependence. Plugging the above into \eqref{eq:spin-alignment-proportional} and \eqref{spin_correlation}, and using the parameterization \eqref{eq:equilibrium-response-input}, we obtain
\begin{equation}
 \left.\rho_{00}-\frac{1}{3}\right|_{\rm rel.\,motion}^{\rm stat}
 =
 \frac{\Xi_{\rm eq}}{18}
 \frac{p_{\rm cm}^2}{M_Q^2}
 \left[1-3(\hat n\!\cdot\!\hat p_{\rm cm})^2\right].
 \label{eq:rho00-relmotion-structure}
\end{equation}
with the minus sign in \eqref{spin_correlation} absorbed by the contraction $\frac{2}{3}\langle B'^2_{stat}\rangle-2\langle(B'\cdot n)^2_{stat}\rangle$ from \eqref{eq:spin-alignment-proportional} to give the factor $\left[1-3(\hat n\!\cdot\!\hat p_{\rm cm})^2\right]$.

\subsection{Anisotropy from shear of QGP}
\label{sec:chap3:shear}

We now consider another source of anisotropic chromo-magnetic field fluctuations, namely the shear
of the QGP. Nonvanishing shear tensor arises in a non-central heavy-ion collision, where the fireball
expands non-uniformly because of the anisotropic initial geometry and the longitudinal expansion.
The shear induced anisotropy in chromomagnetic field fluctuation can be written in TAG as
\begin{equation}
\vfourrev{\d{\cal C}_B^{ij}(q_0,q)}=\frac{1}{d_A}\delta^{ab}\int dtd^3x \vfourrev{e^{iq_0t-i\vec{q}\cdot\vec{x}}}\d_\s\langle B_a^i(t,x)B_b^j(0)\rangle
\end{equation}
Here $\delta_\sigma$ denotes the first-order response to shear.
It is instructive to revert to the equilibrium fluctuation 
\begin{align}
{\cal C}_{B,ab}^{ij}(q_0,q)=\(\frac{1}{2}+n_B(q_0)\)\r_{B,ab}^{ij}(q_0,q).
\end{align}
The parametric dependence of the equilibrium spectral density reads $\r_{B,ab}^{ij}(q_0,q)\sim q^2\frac{m_D^2q_0}{q^5}$, with the factor $q^2$ from conversion between gluon to chromomagnetic field and the remaining factor from gluon spectral density. Taking $q_0\to0$, we have $${\cal C}_{B,ab}^{ij}(0,{\bf q})\sim \frac{{T}m_D^2\d^{ij}}{q^3},$$
which captures the logarithmic sensitivity to IR cutoff when integrated over $q$.
In the presence of shear tensor, both the distribution factor and spectral density receive off-equilibrium corrections. The former does not give rise to spin alignment as the equilibrium spectral density $\r_{B,ab}^{ij}(q_0,q)$ is isotropic. The latter introduces a modification to thermal mass \cite{York:2008rr}
\begin{align}
\d{\cal C}_{B,ab}^{ij}\sim\frac{{T}\d m_D^2}{q^3},
\end{align}
with a tensor structure determined by the shear tensor. This estimate illustrates the logarithmic infrared sensitivity of the linear magnetic contribution. Appendix~\ref{sec_app_shear_derivation} evaluates the linear magnetic contribution in TAG within the adopted HTL treatment and finds logarithmic sensitivity to the infrared cutoff $\mu$. We treat the shear coefficient as a phenomenological input.

To have an order of magnitude estimate of the shear induced chromomagnetic fluctuation, we can look at a closely related quantity below
\begin{align}\label{q0_int}
\d\langle B_a^i(t,0)B_b^j(t,0)\rangle=\int_{q_0,q}\d{\cal C}_{B,ab}^{ij}(q_0,q).
\end{align}
On dimensional ground, we expect
\begin{align}\label{T_relation}
\int_{q_0,q}\d{\cal C}_{B,ab}^{ij}(q_0,q)\sim T \int_q\d{\cal C}_{B,ab}^{ij}(0,q).
\end{align}
Recall the form of energy-momentum tensor for gluon plasma is given by $$T_g^{ij}=-E_i^aE_j^a-B_i^aB_j^a+\frac{1}{2}\d^{ij}({\bf E}_a^2+{\bf B}_a^2),$$
we find the off-diagonal element of \eqref{q0_int} is simply the gluonic contribution to $T^{ij}$ upon tracing over color. The response of off-diagonal $T^{ij}$ to shear tensor is parameterized by shear viscosity $T^{ij}=-\h \s^{ij}$, with
\begin{equation}
\sigma^{ij}
=
\partial^iu^j
+
\partial^ju^i
-
\frac{2}{3}\delta^{ij}\partial_ku^k,
\end{equation}
in local rest frame of the QGP fluid.
Combining with \eqref{T_relation}, we obtain the following estimate
\begin{align}
\int_q\d{\cal C}_{B,ab}^{ij}(0,q)\sim \frac{\h}{T}\s^{ij}
\end{align}
We will use the following parameterization for the shear induced off-diagonal chromomagnetic fluctuation\footnote{By rotation symmetry, we have the general form $\int_{\bm q}\delta {\cal C}_{B}^{ij}(0,\bm q)=A_1\d^{ij}+A_2\s^{ij}$, with only the second term contributing to off-diagonal elements.}
\begin{align}
\int_{\bm q}\delta {\cal C}_{B,ab}^{ij}(0,\bm q)
&=\delta_{ab}\,{\cal B}_{\rm stat}\,\sigma^{ij},
\end{align}
with ${\cal B}_{\rm stat}\sim \h/T$. Plugging the parameterization into \eqref{eq:spin-alignment-proportional} and \eqref{spin_correlation}, we obtain
\begin{align}\label{eq:shear}
\r_{00}-\frac{1}{3}=\frac{1}{3}\frac{\c_s^2}{\t_sC}{\cal B}_{\rm stat}\s^{ij}n_in_j=\frac{1}{3}\d\Xi_{\s}\frac{\s_{nn}}{T}.
\end{align}
We have expressed the above as the product of two dimensionless combinations $\d\Xi_{\s}=\frac{\c_s^2}{\t_sC}{\cal B}_\text{stat}T$ and $\frac{\s^{ij}n_in_j}{T}=\frac{\s_{nn}}{T}$. The minus sign in \eqref{spin_correlation} is canceled by the minus sign from the contraction below
\begin{equation*}
\left(\frac23\d_{ij}-2n_in_j\right)\s^{ij}=-2n_in_j\s^{ij}.
\end{equation*}
Using the same order of magnitude estimate for $\c_s$ and $\t_s$ as in Sec.~\ref{sec:chap3:general}, we obtain $\d\Xi_{\s}\sim \frac{r_{Q\bar{Q}}^4T^3\h}{M_Q^2}\sim\frac{r_{Q\bar{Q}}^4T^3s}{M_Q^2}\frac{\h}{s}$. Here entropy density $s$ is introduced to normalize the shear viscosity.

\section{Application to \texorpdfstring{$\jp$}{J/psi} spin alignment}\label{sec:jpsi}

We now apply the two sources of anisotropic chromomagnetic-field
fluctuations to $\jp$ spin alignment. At LHC energy, the $J/\ps$ is produced predominantly by regeneration. The momentum distribution of $J/\ps$ inherits that of charm quark through coalescence. The momentum distribution of charm quark can be studied by heavy quark transport equation. Studies of the Boltzmann-Fokker-Planck equation in Bjorken flow indicate incomplete equilibration of charm quark, with transverse and longitudinal momentum variances in local fluid frame being larger and smaller than the equilibrium counterpart respectively \cite{Moore:2004tg}. 

For the purpose of illustration, we adopt the following simple momentum distribution for $J/\ps$: vanishing longitudinal momentum variance and equilibrium transverse momentum variances\footnote{If $J/\ps$ is in complete equilibrium with local QGP, a vanishing spin alignment is expected.}.
We will also take Bjorken flow as fluid model for QGP, for which the fluid velocity is given by
\begin{equation}
u^\mu=(\cosh\eta_s,0,0,\sinh\eta_s).
\end{equation}
Here $\h_s$ is the spacetime rapidity defined by
$\eta_s=\frac{1}{2}\ln\frac{t+z}{t-z}$.
On the other hand, $p^\mu$ is parameterized by the momentum rapidity
$y$ as
\begin{align}
p^\m=(M_T\cosh y,p_T\cos\f,p_T\sin\f,M_T\sinh y),
\end{align}
with
$y=\frac{1}{2}\ln\frac{p^0+p^z}{p^0-p^z}$ and
$M_T=\sqrt{M^2+p_T^2}$. If quarkonia were in local equilibrium, we would have the Boltzmann distribution as $M\gg T$
\begin{align}
f=e^{-p^\m u_\m/T},
\end{align}
with $p^\mu u_\mu=M_T\cosh(y-\eta_s)$. Our assumption of vanishing longitudinal momentum variances and equilibrium transverse variance translates to $y=\h_s$ and $f=e^{-M_T/T}$. 
The temperature is taken to be the freezeout temperature $T=150\rm{MeV}$.

\begin{figure}[t]
	\centering
	
\includegraphics[width=0.78\linewidth]{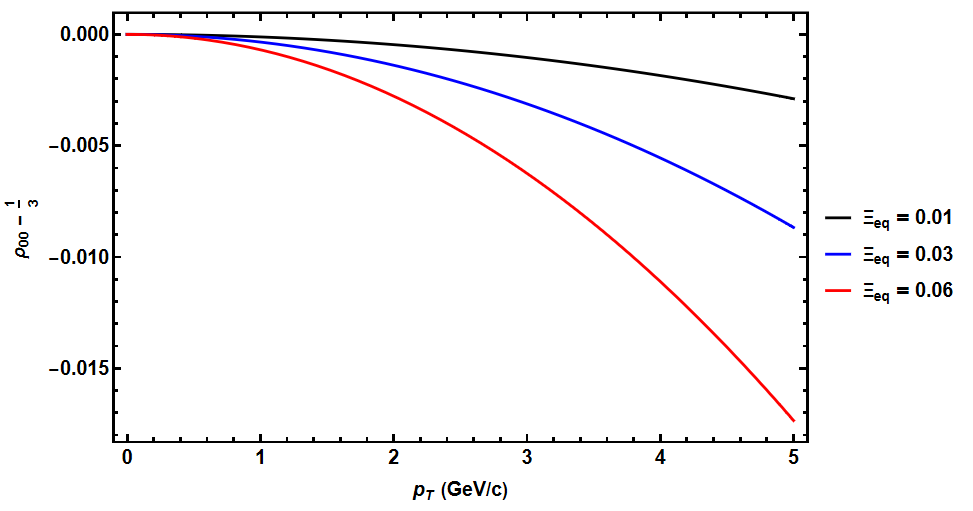}
\caption{
			Relative-motion contribution to the color-singlet $J/\psi$
			spin alignment $\rho_{00}-1/3$ as a function of the transverse
			momentum $p_T^{J/\psi}$ in a Bjorken flow, assuming longitudinal motion locked with local fluid motion and transverse motion reaches equilibrium. The curves evaluate Eq.~
			\eqref{eq:rho00-relmotion-parameterized-static} for
$\Xi_{\rm eq}=0.01,\,0.03,\,0.06$ with $T=150\text{MeV}$ and $M_Q=1.55\text{GeV}$.}
	\label{fig:rho00-relmotion-gaussian}
\end{figure}

Now we discuss the contributions from relative motion and shear respectively. For motion induced spin alignment, we note that with $y=\h_s$, the relative motion is confined to the transverse plane,
\begin{equation}\label{pJpsi}
p_{\mathrm{cm}}^2\simeq \left(p_T^{J/\psi}\right)^2.
\end{equation}
The spin-quantization direction $\hat n$ is chosen perpendicular to the
reaction plane.  After averaging over the azimuthal angle of
$\vec p_T^{\,J/\psi}$ for fixed $p_T$,
\begin{equation}
\left\langle
(\hat n\!\cdot\!\hat p_{\mathrm{cm}})^2
\right\rangle_\phi
=
\frac{1}{2}.
\label{eq:azim-average}
\end{equation}
Plugging \eqref{pJpsi} and \eqref{eq:azim-average} into \eqref{eq:rho00-relmotion-structure}, we obtain
\begin{equation}
 \left.\rho_{00}-\frac{1}{3}\right|_{\rm rel.\,motion}^{\rm stat}
 =-\frac{\Xi_{\rm eq}}{36}
 \frac{(p_T^{J/\psi})^2}{M_Q^2}.
 \label{eq:rho00-relmotion-parameterized-static}
\end{equation}
We take charm quark mass $M_Q=1.55~{\rm GeV}$. Fig.~\ref{fig:rho00-relmotion-gaussian} shows spin alignment of $\jp$ as a function of $p_T$ using three benchmark values
 $\Xi_{\rm eq}=0.01,\, 0.03,\, 0.06$.

\begin{figure}[t]
	\centering
	
\includegraphics[width=0.78\linewidth]{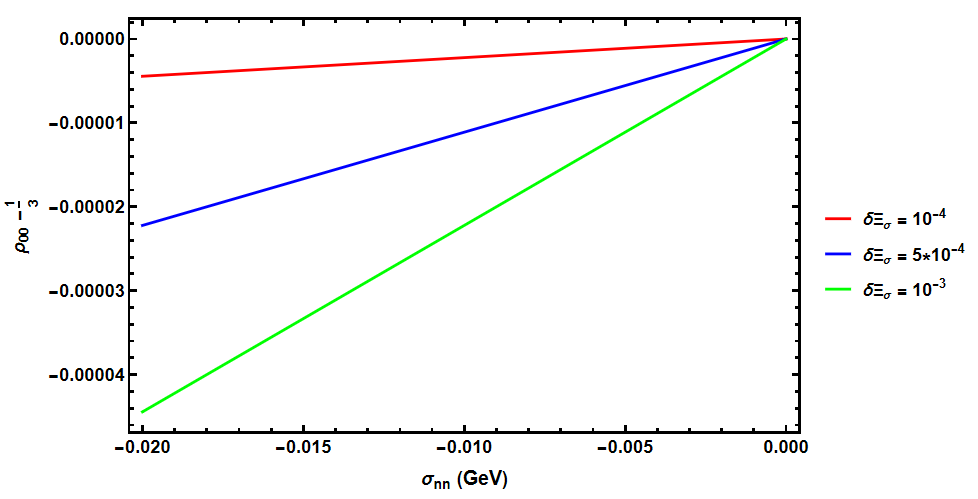}
\caption{
Shear-induced contribution to $J/\psi$ spin alignment
$\rho_{00}-1/3$ as a function of $\sigma^{ij}n_in_j$ in a Bjorken flow. The curves evaluate Eq.~\eqref{eq:shear} for
$\d\Xi_{\s}=10^{-4},\,5\times10^{-4},\,10^{-3}$ with $T=150\text{MeV}$ and $M_Q=1.55\text{GeV}$. %
}
\label{fig:rho00-shear-sigma}
\end{figure}

For shear induced spin alignment, we ignore the relative motion between quarkonium and the QGP. With the spin-quantization direction chosen perpendicular to the reaction plane, $n=\hat y$, the contraction $\s^{ij}n_in_j$ in \eqref{eq:shear} is evaluated as
\begin{equation}
\sigma^{yy}
=
\frac{4}{3}\partial_yu^y
-
\frac{2}{3}\partial_xu^x
-
\frac{2}{3}\partial_zu^z .
\label{eq:sigma-yy-ns}
\end{equation}
In Bjorken flow, only the last term survives to give
$\s^{yy}=-\frac{2}{3\t}$, which gives a negative contribution to spin alignment.  For $\t\simeq10$--$20\,\mathrm{fm}$ this is
$\s^{yy}=-(6.7$--$13.3)\,\mathrm{MeV}$.  To give a reasonable numerical
value to $\d\Xi_{\s}$, we use the form
$\frac{r_{Q\bar{Q}}^4T^3s}{M_Q^2}\frac{\h}{s}$ derived in
Sec.~\ref{sec:chap3:shear}.  The normalized shear viscosity is taken to be
$\h/s\simeq0.08$ \cite{Luzum:2008cw} and the entropy density $s$ is chosen to be free gas of quarks and gluons, with $s\sim T^3$. 
For illustration, we show in Fig.~\ref{fig:rho00-shear-sigma} as a function of $\s_{nn}$ using three benchmark values
$\d\Xi_{\s}=10^{-4},\,5\times10^{-4},\,10^{-3}$. We find shear induced contribution to spin alignment is numerically suppressed compared to motion induced contribution. This is because the shear source is numerically smaller by a factor of $\h/s$ compared to motion source $\d\Xi_{\s}\sim\Xi_{\rm eq}\frac{\h}{s}$ and the shear induced contribution is further multiplied by another numerically small factor $\s_{nn}/T$.

\section{Conclusion and outlook}\label{sec_outlook}

We have studied the spin alignment of quarkonia produced in heavy ion collisions. The spin alignment is found to be a convolution of spin correlation and spin response. The spin correlations can be generated through
chromomagnetic-field fluctuations in the QGP. Anisotropic fluctuations
generically contribute to spin alignment. We considered two
sources of anisotropy: the motion of a quarkonium relative to the QGP,
which makes isotropic fluctuations in the QGP anisotropic in the quarkonium
rest frame, and QGP shear flow, which produces anisotropic viscous corrections to the chromomagnetic field fluctuation. The linear HTL estimates indicate sensitivity to the nonperturbative magnetic sector.
Our gauge-covariant single-mode relaxation model expresses the spin correlation in terms of a chromomagnetic correlator containing an adjoint Wilson line. In TAG the fixed-position temporal Wilson line is the identity. We treat the susceptibility, relaxation time, and field correlators as phenomenological inputs.

We have applied the mechanisms to study the spin alignment of the $\jp$
state in a Bjorken flow. We have studied the contributions to spin alignment from relative motion and shear flow. Both are found to be negative, which is consistent with experimental measurements. The shear induced contribution to spin alignment is found to be numerically suppressed compared to motion induced one.
Further phenomenological studies including realistic production,
regeneration, and dissociation effects are needed for more direct
comparison with experimental data.

We have restricted our study to diagonal elements of the quarkonium
spin density matrix. It is straightforward to extend the analysis to
off-diagonal elements, which are expected to provide more stringent
constraints on mechanisms of spin alignment. It will be interesting to
test the predicted motional effect against future measurements of these
elements.

Finally, the general mechanism is expected to be applicable to $\f$ as
well, with one interesting distinction: the relaxation time of strange quark spin can be comparable to the correlation time of chromomagnetic field. It follows that spin alignment of $\f$ may probe fluctuation of chromomagnetic field at all frequencies instead of the zero frequency limit only. It offers the possibility of connecting spin alignment measurement to real-time nonperturbative dynamics of chromomagnetic field.

\section*{Acknowledgments}
We thank Xiao-Zhi Bai, Defu Hou, Xin-Li Sheng, Yifeng Sun, Qun Wang, Xin-Nian Wang, Di-Lun Yang, Yi Yin and Pengfei Zhuang for stimulating discussions. This work is in part supported by NSFC under Grant Nos 12475148, 12075328.

\appendix

\section{Alternative projection from general spin density matrices}
\label{sec_app_general_density}

The spin-density-matrix structure in Sec.~\ref{sec:density-matrix} can
also be derived by projecting general quark and antiquark spin density
matrices onto the spin-triplet subspace.  Let
\begin{equation}
\rho_Q
=
\frac{1}{2}
\left(1+\mathbf P\cdot\boldsymbol{\sigma}\right),
\qquad
\rho_{\bar Q}
=
\frac{1}{2}
\left(1+\bar{\mathbf P}\cdot\boldsymbol{\sigma}\right).
\label{eq:app-rhoQ-general}
\end{equation}
We define
\begin{equation}
P_{\pm}=P_x\pm iP_y,
\qquad
\bar P_{\pm}=\bar P_x\pm i\bar P_y .
\label{eq:app-Ppm}
\end{equation}
Projecting $\rho_Q\otimes\rho_{\bar Q}$ onto
$|1,1\rangle$, $|1,0\rangle$, and $|1,-1\rangle$ gives
\begin{equation}
\mathcal N
=
\sum_{m=-1}^{1}
\langle 1,m|\rho_Q\otimes\rho_{\bar Q}|1,m\rangle
=
\frac{3+\mathbf P\cdot\bar{\mathbf P}}{4}.
\label{eq:app-triplet-norm}
\end{equation}
The normalized spin-triplet density matrix, in the basis
$|1,1\rangle$, $|1,0\rangle$, $|1,-1\rangle$, is
\begingroup
\small
\setlength{\arraycolsep}{3pt}
\begin{equation}
\rho^{V}
=
\frac{1}{3+\mathbf P\cdot\bar{\mathbf P}}
\begin{pmatrix}
(1+P_z)(1+\bar P_z)
&
\dfrac{(1+P_z)\bar P_-+(1+\bar P_z)P_-}{\sqrt{2}}
&
P_-\bar P_-
\\[8pt]
\dfrac{(1+P_z)\bar P_++(1+\bar P_z)P_+}{\sqrt{2}}
&
1-P_z\bar P_z+P_x\bar P_x+P_y\bar P_y
&
\dfrac{(1-\bar P_z)P_-+(1-P_z)\bar P_-}{\sqrt{2}}
\\[8pt]
P_+\bar P_+
&
\dfrac{(1-\bar P_z)P_++(1-P_z)\bar P_+}{\sqrt{2}}
&
(1-P_z)(1-\bar P_z)
\end{pmatrix}.
\label{eq:app-rhoV-general}
\end{equation}
\endgroup
In particular, for a general spin quantization axis $\hat n$,
\begin{equation}
\rho_{00}
=
\frac{
	1+\mathbf P\cdot\bar{\mathbf P}
	-
	2(\mathbf P\cdot\hat n)
	(\bar{\mathbf P}\cdot\hat n)
}{
	3+\mathbf P\cdot\bar{\mathbf P}
}.
\label{eq:app-rho00-general}
\end{equation}
Keeping only the second-order spin-dependent part gives
\begin{equation}
\rho_{00}-\frac{1}{3}
=
\frac{1}{3}\left[
\frac{2}{3}
\mathbf P\cdot\bar{\mathbf P}
-
2(\mathbf P\cdot\hat n)
(\bar{\mathbf P}\cdot\hat n)
\right]
+O(P^4).
\label{eq:app-rho00-structure}
\end{equation}

\section{Equilibrium chromoelectric and chromomagnetic fluctuations}
\label{sec_app_equilibrium_htl}
    
\rev{This appendix evaluates the zero-frequency equilibrium HTL field correlators in temporal axial gauge, $A_0^a=0$, using field strengths linear in the gauge potential. The fixed-position temporal Wilson line defined in Eq.~\eqref{eq:adjoint-worldline-link} reduces to $U_A=\mathbf{1}$ in this gauge.}

\subsection*{Field-strength correlators and HTL spectral input}

We write $Q=(Q^0,\bm q)$, $q=|\bm q|$, and
$\hat{\bm q}=\bm q/q$, with analogous conventions for other momenta.
Greek Lorentz indices run over $0,1,2,3$, and spatial indices over
$1,2,3$, with spatial contractions taken using $\delta_{ij}$.
We use $g^{\mu\nu}=\operatorname{diag}(-1,1,1,1)$. \vfourrev{This convention applies only to Appendices~\ref{sec_app_equilibrium_htl} and~\ref{sec_app_shear_derivation}; the main text retains the opposite signature, $\operatorname{diag}(1,-1,-1,-1)$.}
\rev{The superscript $a$ on $A_\mu^a$ is a color index.
Here $g_s$ is the gauge coupling, $N_c$ the number of colors,
and $N_f$ the number of quark flavors included in the thermal medium.}

The restriction to linear field strengths is motivated by preliminary power counting in the HTL-resummed soft sector. For characteristic internal momenta $k\sim g_sT$, the statistical gluon propagator scales as $D_{rr}\sim (T/k)k^{-2}$, including the Bose enhancement. Relative to the linear magnetic contraction $k^2D_{rr}$, the contraction of two quadratic field-strength terms $g_sAA$ contains a factor $g_s^2$, an additional loop measure $k^4$, and one additional statistical propagator. Its relative size is therefore $g_s^2k^4D_{rr}/k^2\sim g_s^2T/k\sim g_s$. This counting motivates retaining only the linear field-strength contribution in the present HTL estimate.

In TAG, the field strengths at linear order in the gauge potential are
\begin{align}
B_i^a&\simeq\varepsilon_{ijk}\partial_jA_k^a,
\qquad E_i^a=-\partial_0A_i^a.
\label{eq:app-BE-fields}
\end{align}
The electric relation is exact in TAG. The corresponding  symmetrized field-strength correlators are
\begin{align}
\langle B_iB_l\rangle_Q
&=\varepsilon_{ijk}\varepsilon_{lnm}q_jq_n
\langle A_kA_m\rangle_Q,
\label{eq:BB-general}\\
\langle E_iE_j\rangle_Q
&=(Q^0)^2\langle A_iA_j\rangle_Q.
\label{eq:EE-general}
\end{align}
The common color factor $\delta^{ab}$ is suppressed.

We denote the temperature by $T$ and its inverse by $\beta=1/T$.
The Bose--Einstein distribution is
$n_B(\omega)=[e^{\beta\omega}-1]^{-1}$.
The symmetrized fluctuation--dissipation relation is
\begin{equation}
\langle A_\mu^aA_\nu^b\rangle_Q
=\delta^{ab}w(Q^0)\rho_{\mu\nu}^{{\mathrm{TAG}}}(Q),
\qquad w(\omega)\equiv\frac12+n_B(\omega).
\label{eq:app-fluctuation-dissipation}
\end{equation}
\begin{revision}
The TAG spectral tensor is~\cite{Blaizot:2000TAG}
\begin{equation}
\rho_{ij}^{\mathrm{TAG}}(Q)
=P^T_{ij}\rho_T(Q)+\frac{q_iq_j}{(Q^0)^2}\rho_L(Q),
\qquad \rho_{0\mu}^{\mathrm{TAG}}(Q)=0,\qquad Q^0\ne0.
\label{eq:app-TAG-spectral-tensor}
\end{equation}
Here $P^T_{ij}=\delta_{ij}-\hat q_i\hat q_j$. At HTL order, $\rho_T$ and $\rho_L$ are the same transverse and electric scalar spectral functions used in Coulomb gauge. In TAG, the electric scalar enters the spatial longitudinal component with the factor $q_iq_j/(Q^0)^2$. The magnetic curls eliminate this component, while the electric time derivatives cancel its $(Q^0)^{-2}$ factor. These contractions are performed before taking $Q^0\to0$.

For the zero-frequency limit at fixed $q>0$, the relevant HTL spectral input is the Landau cut~\cite{Bellac:2011kqa},
\begin{equation}
\rho_\alpha^{\rm cut}(Q^0,q)
\equiv2\pi\beta_\alpha(x,q)
\theta\!\left(q^2-(Q^0)^2\right).
\label{eq:app-htl-cut}
\end{equation}
Here $\alpha=T,L$, $x=Q^0/q$, and $m_D^2=g_s^2T^2(N_c+N_f/2)/3$.
The cut weights and their constituent functions are
\end{revision}
\begin{equation}
\begin{aligned}
\beta_T(x,q)
&=\frac{m_D^2x(1-x^2)}{4[a_T(x,q)^2+b_T(x)^2]},
&\beta_L(x,q)&=\frac{m_D^2x}{2[a_L(x,q)^2+b_L(x)^2]},\\
a_T(x,q)
&=q^2(x^2-1)-\frac{m_D^2}{2}
\left[x^2+\frac{x(1-x^2)}{2}L(x)\right],
&b_T(x)&=\frac{\pi m_D^2}{4}x(1-x^2),\\
a_L(x,q)
&=q^2+m_D^2\left[1-\frac{x}{2}L(x)\right],
&b_L(x)&=\frac{\pi m_D^2}{2}x,\\
L(x)&=\ln\frac{1+x}{1-x}, &&0<x<1.
\end{aligned}
\label{eq:htl-cut-functions}
\end{equation}
\subsection*{Static linear terms}
At fixed $q>0$, take $x\to0^+$ and then set $Q^0=0$.  The cut limits are
\begin{equation}
n_B(qx)\beta_T\to\frac{m_D^2T}{4q^5},
\qquad
n_B(qx)\beta_L\to\frac{m_D^2T}{2q(q^2+m_D^2)^2}.
\label{eq:app-static-limit-betaT}
\end{equation}
We denote the ultraviolet cutoff of $q$-integral by $\L$. An infrared cutoff $\mu$ is introduced for the $q$-integral in the magnetic correlator. No infrared cutoff is required for the electric counterpart,
because Debye screening makes the longitudinal static integral finite at
the origin.  Hence
\begin{align}
BB^{(1)}_{\rm stat}(\L,\m)&=\frac{m_D^2T}{6\pi}\ln\frac{\L}{\m},
\label{eq:app-static-BB}\\
EE^{(1)}_{\rm stat}(\L)
&=\frac{m_D^2T}{6\pi}[\Phi(\L)-\Phi(0)],\\
\Phi(q)&=\frac12\left[\ln(q^2+m_D^2)
 +\frac{m_D^2}{q^2+m_D^2}\right].
\label{eq:app-static-EE}
\end{align}
\rev{Within the linear-field-strength HTL calculation, the magnetic correlator is logarithmically infrared sensitive, whereas the electric correlator is infrared finite because of Debye screening.}

\newpage
\section{Shear-induced chromomagnetic fluctuation in HTL approximation}
\label{sec_app_shear_derivation}
\rev{As in Appendix~\ref{sec_app_equilibrium_htl}, we work in TAG, where the fixed-position temporal Wilson line is exactly $U_A=\mathbf{1}$, and retain the linear magnetic field strength. The magnetic part project the propagators onto their transverse components, so the shear calculation below uses the transverse HTL propagators.}

We now compute the first-order shear correction to the static
chromomagnetic fluctuation, using the equilibrium HTL propagators and
infrared analysis of Appendix~\ref{sec_app_equilibrium_htl}.
We first express the shear correction to the fluctuation in terms of the correction to spectral functions. We then give the required shear self-energy and evaluate the linear magnetic contribution.

\subsection*{Spectral correction}

Let $\delta\Pi^{\m\n,</>}$ denote the first-order shear correction to the lesser/greater
self-energy and $D_{\m\n}^{R,A}$ the equilibrium HTL propagators. The resulting correction to lesser/greater propagators are given by \cite{Blaizot:1999xk}
\begin{align}
\d D_{\m\n}^{</>}= -D_{\m\a}^R\d\P^{\a\b</>}D_{\b\n}^A,
\end{align}
which is valid to first order in gradient expansion.
Using the exact relation $D_{\m\n}^{rr}=\frac{1}{2}\(D_{\m\n}^<+D_{\m\n}^>\)$, we have
\begin{align}
\d D_{\m\n}^{rr}=-\frac{1}{2}D_{\m\a}^R\(\d\P^{\a\b<}+\d\P^{\a\b>}\)D_{\b\n}^A.
\end{align}
In equilibrium, we have
\begin{align}
\P^{\a\b<}+\P^{\a\b>}=\(1+2n_B(q_0)\)\r^{\a\b},
\end{align}
with $\r^{\a\b}$ being the spectral density. We assume the tensorial form of the above representation holds out of equilibrium, i.e.
\begin{align}
\d\P^{\a\b<}+\d\P^{\a\b>}=\(O(\s)\)\r^{\a\b}+\(1+2n_B(q_0)\)\d\r^{\a\b}.
\end{align}
The first term inherits equilibrium spectral density thus is manifestly isotropic and thus cannot contribute to spin alignment. We evaluate the second term only using the following exact representation
\begin{align}
\r^{\a\b}=i\(D^{\a\b}_A-D^{\a\b}_R\)=2\text{Im}D^{\a\b}_R,
\end{align}
to obtain
\begin{equation}
\delta\langle A_iA_j\rangle(Q)
=
w(q^0)
\left[
2D^R(Q)\,
\operatorname{Im}\delta\Pi^R(Q)\,
D^A(Q)
\right]_{ij}.
\label{eq:shear-RA-correlator}
\end{equation}
Contraction of Lorentz indices is understood as matrix multiplication in the square bracket. The transverse HTL functions $a_T$ and $b_T$ are given in Eq.~\eqref{eq:htl-cut-functions}. It remains to calculate the shear correction to spectral density.

\subsection*{Shear self-energy input}

We write $C_A=N_c$ for the adjoint quadratic Casimir of $SU(N_c)$.
The gluon shear-response function $\chi_g(p)$ is defined by
$\delta f_g(\bm p)=-\frac12 n_B(p)[1+n_B(p)]\chi_g(p)
\sigma_{ij}\hat p_i\hat p_j$, where $p=|\bm p|$ is also the energy
of a massless hard gluon. Here $\sigma_{ij}$ is the symmetric,
traceless shear tensor in the local rest frame, normalized as
$\sigma_{ij}=\partial_i u_j+\partial_j u_i
-\frac23\delta_{ij}\partial_k u_k$.
This definition fixes the sign and normalization of $\chi_g$;
only the gluon contribution to the shear self-energy is included below.
To first order in the shear gradient, the retarded hard-loop self-energy
derived in Appendix~A of Ref.~\cite{York:2008rr} can be written as
\begin{equation}
\begin{aligned}
\delta\Pi_{\mu\nu}^{R}(Q)
&=\beta\,\delta m_g^2 A_{\mu\nu}^{R}(Q),\\
\delta m_g^2
&=-\frac{g_s^2C_AT}{\pi^2}
\int_0^\infty dp\,p\,n_B(p)[1+n_B(p)]\chi_g(p).
\end{aligned}
\label{eq:shear-self-energy}
\end{equation}
Here $\delta m_g^2$ is the shear-induced correction to the gluon
thermal-mass coefficient. The tensor and frequency dependence is contained in
$A_{\mu\nu}^{R}(Q)$.  With $v^\mu=(1,\hat{\bm v})$ and
$x=Q^0/q$, its angular representation is
\begin{equation}
\begin{aligned}
A_R^{\mu\nu}(Q)
&=\frac{\sigma_{ab}}{2}\int\frac{d\Omega_v}{4\pi}
\left(v_a\delta_{bk}+v_b\delta_{ak}-4v_av_bv_k
+\frac{2}{3}\delta_{ab}v_k\right)
\left(v^\mu g^{k\nu}
-\frac{v^\mu v^\nu q^k}{\bm v\!\cdot\!\bm q-Q^0-i0^+}\right).
\end{aligned}
\label{eq:app-Amunu-angular}
\end{equation}
This form makes the dependence on the symmetric-traceless tensor
$\sigma_{ab}$ explicit.

For the transverse shear tensor,
\begin{align}
A_{km}^{R,T}(Q)&=P^T_{kr}A^R_{rs}(Q)P^T_{sm}\notag\\
&=\frac{\sigma_{ab}}{2}\bigg[
C_1^R(x)P^T_{km}\hat q_{\langle a}\hat q_{b\rangle}
\notag\\[-2pt]
&\hspace{1.3cm}+C_2^R(x)\left(P^T_{ka}P^T_{bm}
 +P^T_{kb}P^T_{am}-\frac23\delta_{ab}P^T_{km}\right)\bigg],
\label{eq:app-YM-transverse-tensor-static}
\end{align}
where $\hat q_{\langle a}\hat q_{b\rangle}=\hat q_a\hat q_b-\delta_{ab}/3$. The dimensionless coefficient functions $C_{1,2}^R(x)$ enter below only through their Landau-cut imaginary parts, which for $|x|<1$ are
\begin{equation}
\operatorname{Im}C_1^R=-\frac{\pi}{4}x(1-x^2)(3-5x^2),
\qquad \operatorname{Im}C_2^R=\frac{\pi}{4}x(1-x^2)^2.
\label{eq:app-shear-C12}
\end{equation}
Substituting this shear input into Eq.~\eqref{eq:BB-general} gives the linear magnetic contribution considered next.

\subsection*{Static linear magnetic contribution}
Contracting the shear-corrected transverse correlator with the linear
magnetic field strengths gives the kernel
\begin{equation}
\begin{aligned}
I_{il}^{(1)}(x)
&\equiv\int\frac{d\Omega_k}{4\pi}
\varepsilon_{ijk}\varepsilon_{lnm}\,k_j k_n [2ImA^{R,T}_{km}(x,\hat{\bm k})]
=k^{2}K^{(1)}(x)\,\sigma_{il}\,,\\
K^{(1)}(x)
&=-\frac{2}{15}\operatorname{Im}C_1^R(x)
-\frac{2}{3}\operatorname{Im}C_2^R(x)
=-\frac{\pi}{15}\,x(1-x^2)\,.
\end{aligned}
\label{eq:app-static-kernel}
\end{equation}
Using
$D_{H,T}^RD_{H,T}^A=1/(a_T^2+b_T^2)$ gives
\begin{align}
{\cal B}_{\rm stat}^{(1)}(\L,\m)&=\frac{\beta\,\delta m_g^2}{2\pi^2}
\int_\m^\L dq\,q^4\lim_{x\to0^+}\left[n_B(qx)\frac{K^{(1)}(x)}{a_T^2+b_T^2}\right]
\notag\\
&=-\frac{\beta\,\delta m_g^2T}{30\pi}\ln\frac{\L}{\m}.
\label{eq:app-static-result}
\end{align}

Within the adopted HTL treatment, the linear shear contribution has logarithmic infrared sensitivity, paralleling the equilibrium magnetic result. The shear coefficient is treated as a phenomenological input in the main text.
\IfFileExists{anisotropy.bib}{
  \bibliographystyle{unsrt}\bibliography{anisotropy.bib}

@article{STAR:2017ckg,
    author = "Adamczyk, L. and others",
    collaboration = "STAR",
    title = "{Global $\Lambda$ hyperon polarization in nuclear collisions: evidence for the most vortical fluid}",
    eprint = "1701.06657",
    archivePrefix = "arXiv",
    primaryClass = "nucl-ex",
    doi = "10.1038/nature23004",
    journal = "Nature",
    volume = "548",
    pages = "62--65",
    year = "2017"
}

@article{STAR:2019erd,
    author = "Adam, Jaroslav and others",
    collaboration = "STAR",
    title = "{Polarization of $\Lambda$ ($\bar{\Lambda}$) hyperons along the beam direction in Au+Au collisions at $\sqrt{s_{_{NN}}}$ = 200 GeV}",
    eprint = "1905.11917",
    archivePrefix = "arXiv",
    primaryClass = "nucl-ex",
    doi = "10.1103/PhysRevLett.123.132301",
    journal = "Phys. Rev. Lett.",
    volume = "123",
    number = "13",
    pages = "132301",
    year = "2019"
}

@article{Becattini:2013fla,
    author = "Becattini, F. and Chandra, V. and Del Zanna, L. and Grossi, E.",
    title = "{Relativistic distribution function for particles with spin at local thermodynamical equilibrium}",
    eprint = "1303.3431",
    archivePrefix = "arXiv",
    primaryClass = "nucl-th",
    doi = "10.1016/j.aop.2013.07.004",
    journal = "Annals Phys.",
    volume = "338",
    pages = "32--49",
    year = "2013"
}

@article{Fang:2016vpj,
    author = "Fang, Ren-hong and Pang, Long-gang and Wang, Qun and Wang, Xin-nian",
    title = "{Polarization of massive fermions in a vortical fluid}",
    eprint = "1604.04036",
    archivePrefix = "arXiv",
    primaryClass = "nucl-th",
    reportNumber = "ICTS-USTC-16-05",
    doi = "10.1103/PhysRevC.94.024904",
    journal = "Phys. Rev. C",
    volume = "94",
    number = "2",
    pages = "024904",
    year = "2016"
}

@article{Liang:2004ph,
    author = "Liang, Zuo-Tang and Wang, Xin-Nian",
    title = "{Globally polarized quark-gluon plasma in non-central A+A collisions}",
    eprint = "nucl-th/0410079",
    archivePrefix = "arXiv",
    reportNumber = "LBNL-56383",
    doi = "10.1103/PhysRevLett.94.102301",
    journal = "Phys. Rev. Lett.",
    volume = "94",
    pages = "102301",
    year = "2005",
    note = "[Erratum: Phys.Rev.Lett. 96, 039901 (2006)]"
}

@article{Liang:2004xn,
    author = "Liang, Zuo-Tang and Wang, Xin-Nian",
    title = "{Spin alignment of vector mesons in non-central A+A collisions}",
    eprint = "nucl-th/0411101",
    archivePrefix = "arXiv",
    reportNumber = "LBNL-56659",
    doi = "10.1016/j.physletb.2005.09.060",
    journal = "Phys. Lett. B",
    volume = "629",
    pages = "20--26",
    year = "2005"
}

@article{Yang:2017sdk,
    author = "Yang, Yang-Guang and Fang, Ren-Hong and Wang, Qun and Wang, Xin-Nian",
    title = "{Quark coalescence model for polarized vector mesons and baryons}",
    eprint = "1711.06008",
    archivePrefix = "arXiv",
    primaryClass = "nucl-th",
    doi = "10.1103/PhysRevC.97.034917",
    journal = "Phys. Rev. C",
    volume = "97",
    number = "3",
    pages = "034917",
    year = "2018"
}

@article{ALICE:2019aid,
    author = "Acharya, Shreyasi and others",
    collaboration = "ALICE",
    title = "{Evidence of Spin-Orbital Angular Momentum Interactions in Relativistic Heavy-Ion Collisions}",
    eprint = "1910.14408",
    archivePrefix = "arXiv",
    primaryClass = "nucl-ex",
    reportNumber = "CERN-EP-2019-251",
    doi = "10.1103/PhysRevLett.125.012301",
    journal = "Phys. Rev. Lett.",
    volume = "125",
    number = "1",
    pages = "012301",
    year = "2020"
}

@article{STAR:2022fan,
    author = "Abdallah, M. S. and others",
    collaboration = "STAR",
    title = "{Pattern of global spin alignment of {\ensuremath{\phi}} and K$^{*0}$ mesons in heavy-ion collisions}",
    eprint = "2204.02302",
    archivePrefix = "arXiv",
    primaryClass = "hep-ph",
    doi = "10.1038/s41586-022-05557-5",
    journal = "Nature",
    volume = "614",
    number = "7947",
    pages = "244--248",
    year = "2023"
}

@article{ALICE:2022dyy,
    author = "Acharya, Shreyasi and others",
    collaboration = "ALICE",
    title = "{Measurement of the J/{\ensuremath{\psi}} Polarization with Respect to the Event Plane in Pb-Pb Collisions at the LHC}",
    eprint = "2204.10171",
    archivePrefix = "arXiv",
    primaryClass = "nucl-ex",
    reportNumber = "CERN-EP-2022-066",
    doi = "10.1103/PhysRevLett.131.042303",
    journal = "Phys. Rev. Lett.",
    volume = "131",
    number = "4",
    pages = "042303",
    year = "2023"
}

@article{Sheng:2022wsy,
    author = "Sheng, Xin-Li and Oliva, Lucia and Liang, Zuo-Tang and Wang, Qun and Wang, Xin-Nian",
    title = "{Spin Alignment of Vector Mesons in Heavy-Ion Collisions}",
    eprint = "2205.15689",
    archivePrefix = "arXiv",
    primaryClass = "nucl-th",
    reportNumber = "USTC-ICTS/PCFT-22-16",
    doi = "10.1103/PhysRevLett.131.042304",
    journal = "Phys. Rev. Lett.",
    volume = "131",
    number = "4",
    pages = "042304",
    year = "2023"
}

@article{Sheng:2023urn,
    author = "Sheng, Xin-Li and Pu, Shi and Wang, Qun",
    title = "{Momentum dependence of the spin alignment of the {\ensuremath{\phi}} meson}",
    eprint = "2308.14038",
    archivePrefix = "arXiv",
    primaryClass = "nucl-th",
    doi = "10.1103/PhysRevC.108.054902",
    journal = "Phys. Rev. C",
    volume = "108",
    number = "5",
    pages = "054902",
    year = "2023"
}

@article{Kumar:2023ghs,
    author = {Kumar, Avdhesh and M{\"u}ller, Berndt and Yang, Di-Lun},
    title = "{Spin alignment of vector mesons by glasma fields}",
    eprint = "2304.04181",
    archivePrefix = "arXiv",
    primaryClass = "nucl-th",
    doi = "10.1103/PhysRevD.108.016020",
    journal = "Phys. Rev. D",
    volume = "108",
    number = "1",
    pages = "016020",
    year = "2023"
}

@article{Yang:2024qpy,
    author = "Yang, Di-Lun",
    title = "{Transverse and longitudinal spin alignment from color fields in heavy ion collisions}",
    eprint = "2411.14822",
    archivePrefix = "arXiv",
    primaryClass = "nucl-th",
    doi = "10.1103/PhysRevD.111.056005",
    journal = "Phys. Rev. D",
    volume = "111",
    number = "5",
    pages = "056005",
    year = "2025"
}

@article{Li:2022vmb,
    author = "Li, Feng and Liu, Shuai Y. F.",
    title = "{Tensor polarization and the dissipative damping of vector meson in QCD Medium}",
    eprint = "2206.11890",
    archivePrefix = "arXiv",
    primaryClass = "nucl-th",
    month = "6",
    year = "2022"
}

@article{Wagner:2022gza,
    author = "Wagner, David and Weickgenannt, Nora and Speranza, Enrico",
    title = "{Generating tensor polarization from shear stress}",
    eprint = "2207.01111",
    archivePrefix = "arXiv",
    primaryClass = "nucl-th",
    doi = "10.1103/PhysRevResearch.5.013187",
    journal = "Phys. Rev. Res.",
    volume = "5",
    number = "1",
    pages = "013187",
    year = "2023"
}

@article{DeMoura:2023jzz,
    author = "De Moura, Paulo Henrique and Goncalves, Kayman J. and Torrieri, Giorgio",
    title = "{Quarkonium spin alignment in a vortical medium}",
    eprint = "2305.02985",
    archivePrefix = "arXiv",
    primaryClass = "hep-ph",
    doi = "10.1103/PhysRevD.108.034032",
    journal = "Phys. Rev. D",
    volume = "108",
    number = "3",
    pages = "034032",
    year = "2023"
}

@article{Xu:2024kdh,
    author = "Xu, Kun and Huang, Mei",
    title = "{Spin alignment of vector mesons induced by local spin density fluctuations}",
    eprint = "2408.06581",
    archivePrefix = "arXiv",
    primaryClass = "hep-ph",
    doi = "10.1103/PhysRevD.110.094034",
    journal = "Phys. Rev. D",
    volume = "110",
    number = "9",
    pages = "094034",
    year = "2024"
}

@article{Sheng:2024kgg,
    author = "Sheng, Xin-Li and Zhao, Yan-Qing and Li, Si-Wen and Becattini, Francesco and Hou, Defu",
    title = "{Holographic spin alignment for vector mesons}",
    eprint = "2403.07522",
    archivePrefix = "arXiv",
    primaryClass = "hep-ph",
    doi = "10.1103/PhysRevD.110.056047",
    journal = "Phys. Rev. D",
    volume = "110",
    number = "5",
    pages = "056047",
    year = "2024"
}

@article{Fu:2023qht,
    author = "Fu, Baochi and Gao, Fei and Liu, Yu-Xin and Song, Huichao",
    title = "{The spin alignment of vector mesons with light front quarks}",
    eprint = "2308.07936",
    archivePrefix = "arXiv",
    primaryClass = "hep-ph",
    doi = "10.1016/j.physletb.2024.138821",
    journal = "Phys. Lett. B",
    volume = "855",
    pages = "138821",
    year = "2024"
}

@article{Zhao:2024ipr,
    author = "Zhao, Yan-Qing and Sheng, Xin-Li and Li, Si-Wen and Hou, Defu",
    title = "{Holographic spin alignment of J/{\ensuremath{\psi}} meson in magnetized plasma}",
    eprint = "2403.07468",
    archivePrefix = "arXiv",
    primaryClass = "hep-ph",
    doi = "10.1007/JHEP08(2024)070",
    journal = "JHEP",
    volume = "08",
    pages = "070",
    year = "2024"
}

@article{Chen:2024hki,
    author = "Chen, Hao-Lei and Fu, Wei-jie and Huang, Xu-Guang and Ma, Guo-Liang",
    title = "{Fluctuations and Correlations of Quark Spin in Hot and Dense QCD Matter}",
    eprint = "2410.20704",
    archivePrefix = "arXiv",
    primaryClass = "hep-ph",
    doi = "10.1103/g1bh-85h4",
    journal = "Phys. Rev. Lett.",
    volume = "135",
    number = "3",
    pages = "032302",
    year = "2025"
}

@article{Chen:2025mrf,
    author = "Chen, Zhishun and Lin, Shu",
    title = "{Polarized dissociation and spin alignment of moving quarkonia in a quark-gluon plasma}",
    eprint = "2501.16596",
    archivePrefix = "arXiv",
    primaryClass = "hep-ph",
    doi = "10.1103/PhysRevD.111.074002",
    journal = "Phys. Rev. D",
    volume = "111",
    number = "7",
    pages = "074002",
    year = "2025"
}

@article{Liang:2025hxw,
    author = "Liang, Yuhao and Lin, Shu",
    title = "{Spin alignment of quarkonia in vortical quark-gluon plasma}",
    eprint = "2502.05866",
    archivePrefix = "arXiv",
    primaryClass = "hep-ph",
    doi = "10.1088/1674-1137/adcc8c",
    journal = "Chin. Phys. C",
    volume = "49",
    number = "8",
    pages = "084105",
    year = "2025"
}

@article{Yan:2025tlx,
    author = "Yan, Guowei and Lin, Shu",
    title = "{Distorted quarkonia and spin alignment}",
    eprint = "2507.22684",
    archivePrefix = "arXiv",
    primaryClass = "hep-ph",
    doi = "10.1103/jg98-9vx9",
    journal = "Phys. Rev. D",
    volume = "113",
    number = "5",
    pages = "054045",
    year = "2026"
}

@article{Zhu:2025rdj,
    author = "Zhu, Xin-Nan and Sheng, Xin-Li and Hou, Defu",
    title = "{Production of K+K- pairs through the decay of {\ensuremath{\phi}} mesons}",
    eprint = "2503.23919",
    archivePrefix = "arXiv",
    primaryClass = "hep-ph",
    doi = "10.1103/7s6h-r457",
    journal = "Phys. Rev. D",
    volume = "112",
    number = "5",
    pages = "056011",
    year = "2025"
}

@article{Lv:2024uev,
    author = "Lv, Ji-peng and Yu, Zi-han and Liang, Zuo-tang and Wang, Qun and Wang, Xin-Nian",
    title = "{Global quark spin correlations in relativistic heavy ion collisions}",
    eprint = "2402.13721",
    archivePrefix = "arXiv",
    primaryClass = "hep-ph",
    doi = "10.1103/PhysRevD.109.114003",
    journal = "Phys. Rev. D",
    volume = "109",
    number = "11",
    pages = "114003",
    year = "2024"
}

@article{Oliva:2026wbo,
    author = "Oliva, Lucia and Wang, Qun and Wang, Xin-Nian",
    title = "{Quark spin correlation inside hyperons}",
    eprint = "2603.10427",
    archivePrefix = "arXiv",
    primaryClass = "hep-ph",
    doi = "10.1103/h88t-phwj",
    journal = "Phys. Rev. C",
    volume = "114",
    number = "1",
    pages = "014914",
    year = "2026"
}

@article{STAR:2025njp,
    author = "Aboona, B. E. and others",
    collaboration = "STAR",
    title = "{Measuring spin correlation between quarks during QCD confinement}",
    eprint = "2506.05499",
    archivePrefix = "arXiv",
    primaryClass = "hep-ex",
    doi = "10.1038/s41586-025-09920-0",
    journal = "Nature",
    volume = "650",
    number = "8100",
    pages = "65--71",
    year = "2026"
}

@article{Yang:2024ejk,
    author = "Yang, Di-Lun and Yao, Xiaojun",
    title = "{Quarkonium polarization in medium from open quantum systems and chromomagnetic correlators}",
    eprint = "2405.20280",
    archivePrefix = "arXiv",
    primaryClass = "hep-ph",
    reportNumber = "IQuS@UW-21-079",
    doi = "10.1103/PhysRevD.110.074037",
    journal = "Phys. Rev. D",
    volume = "110",
    number = "7",
    pages = "074037",
    year = "2024"
}

@article{Hongo:2022izs,
    author = "Hongo, Masaru and Huang, Xu-Guang and Kaminski, Matthias and Stephanov, Mikhail and Yee, Ho-Ung",
    title = "{Spin relaxation rate for heavy quarks in weakly coupled QCD plasma}",
    eprint = "2201.12390",
    archivePrefix = "arXiv",
    primaryClass = "hep-th",
    reportNumber = "RIKEN-iTHEMS-Report-22",
    doi = "10.1007/JHEP08(2022)263",
    journal = "JHEP",
    volume = "08",
    pages = "263",
    year = "2022"
}

@book{Bellac:2011kqa,
    author = "Bellac, Michel Le",
    title = "{Thermal Field Theory}",
    doi = "10.1017/CBO9780511721700",
    isbn = "978-0-511-88506-8, 978-0-521-65477-7",
    publisher = "Cambridge University Press",
    series = "Cambridge Monographs on Mathematical Physics",
    month = "3",
    year = "2011"
}

@article{York:2008rr,
    author = "York, Mark Abraao and Moore, Guy D.",
    title = "{Second order hydrodynamic coefficients from kinetic theory}",
    eprint = "0811.0729",
    archivePrefix = "arXiv",
    primaryClass = "hep-ph",
    doi = "10.1103/PhysRevD.79.054011",
    journal = "Phys. Rev. D",
    volume = "79",
    pages = "054011",
    year = "2009"
}

@article{Moore:2004tg,
    author = "Moore, Guy D. and Teaney, Derek",
    title = "{How much do heavy quarks thermalize in a heavy ion collision?}",
    eprint = "hep-ph/0412346",
    archivePrefix = "arXiv",
    doi = "10.1103/PhysRevC.71.064904",
    journal = "Phys. Rev. C",
    volume = "71",
    pages = "064904",
    year = "2005"
}

@article{Luzum:2008cw,
    author = "Luzum, Matthew and Romatschke, Paul",
    title = "{Conformal Relativistic Viscous Hydrodynamics: Applications to RHIC results at s(NN)**(1/2) = 200-GeV}",
    eprint = "0804.4015",
    archivePrefix = "arXiv",
    primaryClass = "nucl-th",
    reportNumber = "INT-08-07, NT-UW-08-10",
    doi = "10.1103/PhysRevC.78.034915",
    journal = "Phys. Rev. C",
    volume = "78",
    pages = "034915",
    year = "2008",
    note = "[Erratum: Phys.Rev.C 79, 039903 (2009)]"
}

@misc{Zhou:2025spin,
    author = {Zhou, Jie and Zhao, Ying Shan and Sun, Yifeng},
    title = {Covariant equations of motion of massive spinning particles in a background {Yang--Mills} field},
    howpublished = {arXiv:2511.20083 [nucl-th]},
    eprint = {2511.20083},
    archivePrefix = {arXiv},
    primaryClass = {nucl-th},
    year = {2025},
    note = {Version 2, revised 9 June 2026}
}

@article{Blaizot:1999xk,
	author = "Blaizot, Jean-Paul and Iancu, Edmond",
	title = "{A Boltzmann equation for the QCD plasma}",
	eprint = "hep-ph/9903389",
	archivePrefix = "arXiv",
	reportNumber = "SACLAY-SPH-T-99-026, CERN-TH-99-71",
	doi = "10.1016/S0550-3213(99)00341-7",
	journal = "Nucl. Phys. B",
	volume = "557",
	pages = "183--236",
	year = "1999"
}

@article{Blaizot:2000TAG,
    author = {Blaizot, Jean-Paul and Iancu, Edmond and Rebhan, Anton},
    title = {Approximately self-consistent resummations for the thermodynamics of the quark-gluon plasma. I. Entropy and density},
    journal = {Phys. Rev. D},
    volume = {63},
    pages = {065003},
    year = {2001},
    doi = {10.1103/PhysRevD.63.065003},
    eprint = {hep-ph/0005003},
    archivePrefix = {arXiv}
}

@article{Zhang:2025freezeout,
    author = {Zhang, Zhong-Hua and Huang, Xu-Guang},
    title = {Tensor spin polarization induced by curved freeze-out hypersurface},
    journal = {Phys. Rev. D},
    volume = {113},
    pages = {116024},
    year = {2026},
    doi = {10.1103/p2mw-b8kf},
    eprint = {2509.20200},
    archivePrefix = {arXiv},
    primaryClass = {hep-ph}
}

@article{Dong:2021Debye,
    author = {Dong, Lihua and Guo, Yun and Islam, Ajaharul and Strickland, Michael},
    title = {Effective Debye screening mass in an anisotropic quark gluon plasma},
    journal = {Phys. Rev. D},
    volume = {104},
    pages = {096017},
    year = {2021},
    doi = {10.1103/PhysRevD.104.096017},
    eprint = {2109.01284},
    archivePrefix = {arXiv},
    primaryClass = {hep-ph},
    note = {\href{https://arxiv.org/abs/2109.01284}{arXiv:2109.01284}}
}
}{}

\end{CJK}

\end{document}